\documentclass[a4paper]{article}

\usepackage{amsfonts}

\def\be{\begin{equation}}
\def\ee{\end{equation}}
\def\bea{\begin{eqnarray}}
\def\eea{\end{eqnarray}}
\def\({\left(}
\def\){\right)}
\def\<{\left<}
\def\>{\right>}

\def\be{\begin{equation}}
\def\ee{\end{equation}}
\def\bea{\begin{eqnarray}}
\def\eea{\end{eqnarray}}
\def\ben{\begin{eqnarray}}
\def\een{\end{eqnarray}}
\def\({\left(}
\def\){\right)}
\def\<{\left<}
\def\>{\right>}

\def\[{\left[}
\def\]{\right]}

\def\+{\bar}
\def\mb{\mathbb}

\def\t{\tilde}

\def\N{{\cal{N}}}

\begin{document}
\setlength{\unitlength}{1mm}

\pagestyle{empty}
\vskip-10pt
\vskip-10pt
\hfill 
\begin{center}
\vskip 3truecm
{\large \bf
M5 brane from mass deformed BLG theory}\\ 
\vskip 2truecm
{\large \bf
Andreas Gustavsson\footnote{a.r.gustavsson@swipnet.se}}\\
\vskip 1truecm
{\it  Center for quantum spacetime (CQUeST), Sogang University, Seoul 121-742, Korea}\\
and\\
\it{School of Physics \& Astronomy, Seoul National University, Seoul 151-747 Korea}
\end{center}
\vskip 2truecm
{\abstract{We study small fluctuations around a BPS three-sphere vacuum solution of mass deformed BLG theory. We realize the BLG theory by a Nambu bracket and find a maximally supersymmetric lagrangian for the fluctuation fields corresponding to a single M5 brane on $\mb{R}^{1,2}\times S^3$.}}

\vfill 
\vskip4pt
\eject
\pagestyle{plain}

\section{Introduction}
The low energy effective theory living on finitely many coincident M2 branes probing the orbifold singularity $\mb{R}^8/\mb{Z}_k$ was found in \cite{ABJM}. It is a Chern-Simons theory with gauge group $U(N)\times U(N)$ coupled to matter fields with manifest $\N=6$ supersymmetry and $SU(4) \times U(1)$ R symmetry. For a given gauge group, the only free parameter is the integer valued Chern-Simons level $k$. For levels $k=1,2$ the theory has enhanced $OSp(8|4)$ maximal superconformal symmetry \cite{ABJM,Gustavsson:2009pm,Kwon:2009ar,Benna:2009xd,Klebanov:2008vq}.

Subsequently a larger class of $\N=6$ superconformal theories were found for various gauge groups \cite{Hosomichi:2008jb}. We will refer to any $\N=6$ supersymmetric Chern-Simons-matter theory  as ABJM theory. It is unclear to me whether all these theories correspond to M2 branes probing an $\mb{R}^8/\mb{Z}_k$ singularity. In any case, for $k=1,2$ all these ABJM theories get enhanced $OSp(8|4)$ superconformal symmetry. 

ABJM theories can also be formulated using a particular class of three-algebras \cite{BL} called hermitian three-algebras. Another type of three-algebra has been found for the $\N=5$ supersymmetric theories \cite{Chen:2009cw}.

The smallest non-trivial ABJM gauge group is $SO(4)$. For this choice of gauge group, the ABJM lagrangian can be recast in a form that is manifestly $OSp(8|4)$ invariant, which is then the BLG lagrangian \cite{BLG} up to a triality of $SO(8)$ R-symmetry indices. 

There are mass deformations of BLG and ABJM theories \cite{Lambert:2009qw,Hosomichi:2008qk,Hosomichi:2008jb} (older works on mass deformed M2 brane theory are from gravity point of view \cite{Bena:2000zb} and from matrix theory point of view \cite{SheikhJabbari:2005mf}) that preserve all the manifest supersymmetries. For ABJM theories this means $\N=6$ supersymmetry. However the $SO(6)$ R symmetry is broken by the mass deformation to $SO(4) \times SO(2)$. For BLG theory the mass deformation preserves $\N=8$ supersymmetry and breaks the $SO(8)$ R symmetry down to $SO(4)\times SO(4)$. It is plausible that also the above mentioned mass deformed ABJM theories will get enhanced $\N=8$ supersymmetry for levels $k=1,2$, along with an enhanced $SO(4) \times SO(4)$ R symmetry. 

For levels $k=1,2$ then, we can in mass deformed ABJM and BLG theories, find a vacuum solution which preserves $\N=8$ supersymmetry. Thus
\bea
\mb{R}^{1,2} \times \frac{S^3_{\mbox{fuzzy}}}{\mb{Z}_k}.
\eea
The fuzzy three-sphere is described by four matrices $G^i$ of a certain size $N\times N$ \cite{Guralnik:2000pb}. This construction generalizes the fuzzy two-sphere construction in \cite{Madore:1991bw}. In the large $N$ limit we can map these matrices to the embedding functions $T^i$ of a classical three-sphere. These obey the three-algebra and three-sphere constraint
\bea
\{T^i,T^j,T^k\} &=& \frac{1}{R}\epsilon^{ijkl} T^l,\cr
T^iT^i &=& R^2\label{two}
\eea
respectively, for a three-sphere of radius $R$. The curly three-bracket is the Nambu bracket as defined in Eq (\ref{Nambu}). (Our general definition is in Eq (\ref{Nambugeneral})). Since 
\bea
K^{ij} &=& \{T^i,T^j,\cdot\}
\eea
are nothing but the six Killing vectors on the three-sphere generating the rotation group $SO(4)$, we have a realization of the $SO(4)$ three-algebra which is the smallest non-trivial three-algebra, and in fact the only possible three-algebra of finite dimension (if we assume a few requirements which are all very natural from physics point of view). However there is an infinite dimensional extension of the $SO(4)$ three-algebra, which is generated by any function on $S^3$ which has a Taylor series expansion 
\bea
f(T^i) = \sum_{k=1}^{\infty} f_{i_1...i_k} T^{i_1} ... T^{i_k}.
\eea
Due to the three-sphere constraint on $T^i T^i$ we only need to consider traceless symmetric tensors $f_{i_1...i_k}$. We could now consider new three-algebra generators 
\bea
T^{i_1...i_k} &=& T^{i_1} \cdots T^{i_k}
\eea
and we find that all these generate an infinite dimensional three-algebra.

In line with these considerations it is natural to also expect that ABJM theory with gauge group $U(N)\times U(N)$, in the large $N$ limit can be mapped into BLG theory which is realized by a Nambu three-bracket on $S^3/{\mb{Z}_k}$ which should be viewed as an $S^1/\mb{Z}_k$ bundle over $S^2$, so that in particular the large $k$ limit is $S^2$ \cite{Nastase:2009ny}. As an aside, since BLG theory is maximally supersymmetric for any $k$, this means that we should find supersymmetry enhancement in ABJM theory in the large $N$ limit for any level $k$.

In this paper we will only study BLG theory with a Nambu bracket on $S^3$. As argued in the paragraph above, this seems to correspond to taking $k=1$ and $N=\infty$ in ABJM theory.

\subsection*{Fluctuation analysis}
In the spirit of \cite{Iso:2001mg,Nastase:2009ny}, we will obtain the induced theory of small fluctuations about the maximally supersymmetric three-sphere vacuum solution of BLG theory. If we temporarily let $X$ collectively denote all the fields in BLG theory, then we expand the mass deformed BLG lagrangian in small fluctuations around the vacuum. We thus write $X = T+\delta X$ where $T$ is the vacuum configuration, and expand the lagrangian as
\bea
{\cal{L}}(X) &=& {\cal{L}}(T) + \delta X \frac{\delta {\cal{L}}}{\delta X} + \frac{1}{2}(\delta X)^2 \frac{\delta {\cal{L}}}{\delta X \delta X} + ...
\eea
All derivatives are evaluated at $T$. If $T$ is a static supersymmetric vacuum, then the lagrangian is minus the hamiltonian and this is minimized at the supersymmetric vacuum. Hence the first order derivatives all vanish and we are left with
\bea
{\cal{L}}(X) &=& {\cal{L}}(T) + \frac{1}{2}(\delta X)^2 \frac{\delta {\cal{L}}}{\delta X \delta X} + ....
\eea
In a static supersymmetric vacuum we have ${\cal{L}}(T) = 0$ and we need not write out the zeroth order term ${\cal{L}}(T)$. However, ${\cal{L}}(T) (=0)$ is invariant only under the unbroken supersymmetries. If we do not write out the term ${\cal{L}}(T)$ (since it is zero anyway), then it looks like the supersymmetry variation of the full action can be found be just computing the supersymmetry variation of the second order term. This is wrong. Zero need not be invariant under a variation. We may consider a vacuum in which $\psi = 0$. This does not mean that the supersymmetry variation of $\psi$ must also be $=0$. In fact the condition for $\delta \psi = 0$ defines in this case the unbroken supersymmetries. Since the higher order terms must cancel the supersymmetry variation of the zeroth order term (because the sum is equal to ${\cal{L}}(X)$ which is the maximally supersymmetric lagrangian), we see that the higher order terms can not be invariant under the broken supersymmetries either. On the other hand, the higher order terms must be invariant under the unbroken supersymmetries since the total lagrangian is invariant, as well as the zeroth order term \cite{Park}.

Previous work on relating BLG theory with a Nambu three-bracket to M5 brane can be found in \cite{Ho:2008nn,Pasti:2009xc,Bonelli:2008kh,Park:2008qe,Bandos:2008fr}. In \cite{Bandos:2008fr} the Carrollian limit of BLG theory (where the speed of light goes to zero) with a Nambu bracket was derived from a single M5 in an infinite tension limit.

Since many calculations in our paper are the same as those in \cite{Ho:2008nn}, we should contrast those calculations with ours. In \cite{Ho:2008nn} the BLG theory is expanded about some background $T$ in which three scalar fields acquire a non vanishing vev. This background does not provide any scale parameter which can be used to perform a systematic fluctuation analysis. Instead the coupling constant $1/k$ in BLG theory must be used as expansion parameter. This means that the strong couling regime  of BLG theory can not be treated. The connection between the background $T$ and the internal three-manifold on which the Nambu three-bracket is to be defined, is left unspecified. Naively the background $T$ in these papers appears to be non-supersymmetric. However BLG theory also has a shift symmetry of the fermion. By breaking this shift symmetry one can render the background invariant under modified BLG supersymmetry variations where one has added a constant shift to the variation of the fermion \cite{Park}. One may then restore the shift symmetry of the fermion (albeit the fermion now is located at a shifted value) in BLG theory and find that this shift symmetry transmutes into a gauge symmetry (a constant shift proportional to the volume form on the three-manifold) that acts on the background three-form gauge potential $C$, from the M5 brane brane point of view. For this approach to work one must also specify some condition on the supersymmetry parameter living on the three-manifold. Perhaps this approach can be consistent on a flat three-torus appropriately embedded in transverse space, on which we may have a constant spinor. The connection between the shift symmetry of the BLG fermion and the gauge variation of the constant background C-field could be interesting and worth further study. We note that the M2 brane also couples electrically to $C$ but this field does not seem to alter the BLG theory as long as $C$ is constant, however its field strength $dC$ has the effect of mass deforming BLG theory \cite{Lambert:2009qw}. 

In this paper we instead follow the approach of \cite{Iso:2001mg,Nastase:2009ny}. We expand about a maximally supersymmetric three-sphere vacuum solution in mass deformed BLG theory. This background provides us with a mass parameter that we can use to quantify the smallness of our fluctuation fields. Hence we can have a small value on $k$ and still have a sensible fluctuation expansion by having a small mass parameter as expansion parameter. Since the background does not break any supersymmetry we find a maximally supersymmetric M5 brane theory on a three-sphere.

The theory of a single M5 brane is subtle due to the selfdual three-form. On a topologically non-trivial space-time one can find several different quantum theories of the selfdual three-form \cite{Witten:1996hc}. Consequently the lagrangian of the selfdual three-form can not be unique, but there must be one lagrangian for each such theory. It seems plausible that this is related to the fact that one can not write down a manifestly covariant lagrangian \cite{Perry:1996mk}. Then it could be that by making a large diffeormorphism (a diffeomorphism not continuously connected to the identity) one transforms one lagrangian into another.

\section{Infinite-dimensional mass-deformed BLG theory}
Our starting point will be BLG theory, realized by a Nambu bracket on some internal three-manifold $M_3$. A priori, if we let $\theta^{\alpha}$ denote some local coordinates on $M_3$, $X^I = X^I(\theta)$ denote the scalar fields and $T^I = T^I(\theta)$ denote their vacuum expectation values, there appears to be a many alternative ways to define such a three-manifold $M_3$:
\begin{enumerate}
\item we may define $M_3$ as an auxiliary three-manifold that is not at all related to the scalar fields $X^I$ nor their vacuum expectation values. In this case the metric on $M_3$ must also be auxiliary and have nothing to do with the scalar fields.
\item we may define $M_3$ as the three-manifold that is embedded into transverse space as $\theta^{\alpha} \mapsto X^I(\theta)$. 
\item we may define $M_3$ as the three-manifold that is embedded into transverse space as $\theta^{\alpha} \mapsto T^I(\theta)$. 
\end{enumerate}
Alternative 1 is unphysical in the since that any three-manifold in eleven-dimensional space-time should be associated to some field configuration $X^I$. If we would ignore this and try to define the Nambu bracket using covariant derivatives (or even just using ordinary derivatives) with respect to the auxiliary metric on this three-manifold, we can obtain a supersymmetric BLG theory using such a three-bracket only if the supersymmetry parameter can be taken to be covariantly constant. We explain why this is the case in the discussion around Eq (\ref{more}) below. Since the supersymmetry parameter carries an R symmetry spinor index in addition to its three-dimensional space-time spinor index, this means that $M_3$ must be flat in order to allow for covariantly constant spinors. This flatness constraint can be understood from the requirement $D_{\alpha}\epsilon=0$ upon commuting two covariant derivatives acting on the supersymmetry parameter $\epsilon$. This flatness condition is a too strong constraint on $M_3$. Alternative 2 implies a dynamically defined three-bracket which varies itself under supersymmetry variations of $X^I$. This then will not give a supersymmetric BLG theory as the variation of the three-bracket itself will contribute with additional, unwanted, terms. Alternative 3 is the only alternative that can give a supersymmetric BLG theory without constraining $M_3$ to be flat.

It is alternative 3 that uniquely should correspond to the large $N$ limit of ABJM theory with gauge group $U(N)\times U(N)$. The three-sphere should be the large $N$ limit of the fuzzy three-sphere vacuum solution of mass deformed ABJM theory at level $k=1$. These things are well-understood for the fuzzy two-sphere. The fuzzy two-sphere is defined in terms of generators of $SU(2)$ in some $N+1$ dimensional representation say, where $N$ can be any positive integer. In the large $N$ limit we can map these $SU(2)$ generators into the three Killing vectors $K^i$ on $S^2$. These Killing vectors in turn, can be expressed in terms of the Poisson bracket as
\bea
K^i &=& \{T^i,\cdot\}
\eea
where $T^i$ describes the embedding of the two-sphere into ${\mb{R}}^3$. The Poisson bracket is defined using the metric on the two-sphere. These  $K^i$ obey the SU(2) algebra as a consequence of the Jacobi identity. 

The obvious generalization to the three-sphere is that in the large $N$ limit, the fuzzy three-sphere generators are mapped into coordinates $T^i$ that describe the embedding of the three-sphere into ${\mb{R}}^4$. The six Killing vectors on the three-sphere are
\bea
K^{ij} &=& \{T^i,T^j,\cdot\}
\eea
The Nambu bracket is defined using the metric on the three-sphere. The Killing vectors then generate the $SO(4)$ Lie algebra as a consequence of the fundamental identity and Eq (\ref{two}). We note that even though the definition of the discrete three-bracket and matrix three-algebra generators in ABJM theory has been obtained explicitly \cite{BL}, it is more subtle to understand the $SO(8)$ R symmetry in terms of this three-bracket at level $k=1$. This necessarily requires proper understanding of monopole operators. Using these monopole operators we have found that the ABJM three-bracket becomes essentially totally antisymmetric \cite{Gustavsson:2009pm}. This is a promising property if it is to be mapped into a totally antisymmetric Nambu bracket in the large $N$ limit. But due to the complication of having to involve monopole operators, we have not yet obtained a rigorous way of taking the large $N$ limit of the ABJM theory three-bracket. Taking this large $N$ limit in a rigorous will be very interesting and we believe that this will eventually lead to an understanding of the theory of multiple M5 branes. 

In order to allow for other supersymmetric vacua apart from the three-sphere of mass deformed theory, we will in the rest of this section assume a more generic vacuum three-manifold and denote its embedding in transverse space as $\theta^{\alpha} \mapsto T^I(\theta)$. We denote Minkowski coordinates on $\mb{R}^{1,2}$ as $x^{\mu}$. We introduce normal coordinates $x^A$ ($A=1,...,5$) to $M_3$ in $\mb{R}^8$. We consider the change of coordinates in $\mb{R}^8$
\bea
(\theta^{\alpha},x^A) & \mapsto & x^I = x^I(\theta,x^A).
\eea
The submanifold $M_3$ is located at constant values of $x^A$, that we can set to $x^A = 0$ so that
\bea
T^I(\theta) &=& x^I(\theta,x^A=0)
\eea
defines a parametrization of $M_3$. The induced metric on $M_3$ is given by
\bea
g_{\alpha\beta} &=& \partial_{\alpha}T^I \partial_{\beta}T^I.
\eea
We will also need
\bea
g_{AB} &=& \partial_A T^I \partial_B T^I,\cr
g_{A\alpha} &=& 0
\eea
on $M_3$. We define the Nambu bracket of three scalar functions $f$, $g$ and $h$ on $M_3$ as
\bea
\{f,g,h\} &=& \sqrt{g}\epsilon^{\alpha\beta\gamma}\partial_{\alpha}f\partial_{\beta}g\partial_{\gamma}h\cr
&=& *(df\wedge dg \wedge dh).\label{Nambu}
\eea
We use the convention that 
\bea
\epsilon_{123} &=& 1
\eea
and all indices are rised by the inverse metric $g^{\alpha\beta}$. Here the star $*$ denotes the Hodge dual on $M_3$. 

It is very important to stress that the Nambu bracket is calculated with respect to a background metric associated to a vacuum state. Hence the supersymmetry variation of the metric is zero. In that sense, BLG theory with a Nambu bracket, appears to make no sense unless one specifies a non-vanishing vacuum field configuration $X^I = T^I$. It is true that the Nambu bracket always satisfies the fundamental identity on any auxiliary three-manifold. However this is not enough to insure supersymmetry. When checking closure of supersymmetry one needs to make a second supersymmetry variation of the fermion. This will involve a term
\bea
\{\bar{\epsilon}\Gamma^I \psi,X^J,X^K\}.\label{more}
\eea
In order to secure on-shell closure one needs to be able to rewrite this as
\bea
\bar{\epsilon}\Gamma^I \{\psi,X^J,X^K\}.
\eea
The same type of problem arises when checking supersymmetry of the BLG action. In both cases one has to be able to freely move out the supersymmetry parameter outside the Nambu bracket. On a generic three-manifold it is not possible to have a covariantly constant spinor. This means we can not obtain a supersymmetric BLG theory if we define our Nambu bracket on a generic three-manifold despite the Nambu bracket obeys the fundamental identity.

We introduce a complete set of functions $T^a(\theta)$ on $M_3$ that will be our generators for the infinite-dimensional three-algebra. We expand the matter fields as
\bea
X^I(x,\theta) &=& X^I_a(x) T^a(\theta),\cr
\psi(x,\theta) &=& \psi_a(x) T^a(\theta)
\eea
We define the gauge covariant derivative as
\bea
D_{\mu} X^I &=& \partial_{\mu} X^I - A_{\mu,ab}\{T^a,T^b,X^I\}.
\eea
We use eleven-dimensional spinor notation since we wish to treat $\mb{R}^{1,2}$ and $M_3$ on the same footing, and eventually identify $\mb{R}^{1,2} \times M_3$ as the world-volume of M5 brane. The supersymmetry parameter $\epsilon$ and spinor field $\psi$ are subject to the chirality conditions
\bea
\t \Gamma \epsilon &=& \epsilon,\cr
\t \Gamma \psi &=& -\psi
\eea
where 
\bea
\t \Gamma &=& \Gamma_{012}.
\eea
We have the following ${\cal{N}}=8$ supersymmetry variations 
\bea
\delta X^I &=& i \bar{\epsilon} \Gamma^I \psi\cr
\delta \psi &=& \Gamma^{\mu}\Gamma_I \epsilon D_{\mu} X^I - \frac{1}{6} \Gamma_{IJK} \epsilon \{X^I,X^J,X^K\},\cr
\delta A_{\mu, ab} &=& i\bar{\epsilon} \Gamma_{\mu} \Gamma_I X^I_{[a} \psi_{b]}.\label{BLG}
\eea
closing on-shell,
\bea
\Gamma^{\mu} D_{\mu} \psi + \frac{1}{2} \Gamma_{IJ}\{X^I,X^J,\psi\} &=& 0,\label{BLG0}\\
D^2 X_I - \frac{i}{2} \{\bar{\psi}, \Gamma_{IJ} X^J, \psi\} - \frac{\partial V}{\partial X^I} &=& 0,\label{BLG1}\\
F_{\mu\nu,ab} + \epsilon_{\mu\nu\lambda} \(X_{I[a} D^{\lambda} X^I_{b]} + \frac{i}{2} \bar{\psi}_{[a}\Gamma^{\lambda} \psi_{b]}\) &=& 0\label{BLG2}.
\eea
Here
\bea
V = \frac{1}{12} \<\{X^I,X^J,X^K\},\{X^I,X^J,X^K\}\>
\eea
and the trace form is defined as
\bea
\<F,G\> &=& \int d^3 \theta \sqrt{g} F G.
\eea
The matter part of the lagrangian density is
\bea
{\cal{L}}_{matter} &=&-\frac{1}{2} \<D_{\mu} X^I,D^{\mu} X^I\> - V \cr
&&+\frac{i}{2} \<\bar{\psi},\Gamma^{\mu}D_{\mu} \psi\> + \frac{i}{4} \<\bar{\psi},\{\Gamma^{IJ}\psi,X^I,X^J\}\>.
\eea
The gauge field part is given by the Chern-Simons term,
\bea
{\cal{L}}_{CS} &=& \frac{1}{2} \epsilon^{\mu\nu\lambda} A_{\mu,ab}\partial_{\nu}A_{\lambda,cd}\<T^a,\{T^b,T^c,T^d\}\> + ...
\eea
The cubic interaction term (denoted by the ellipses) in the Chern-Simons action will not be of any interest to us in this paper.

\subsection*{Mass deformation}
There is a mass deformation of BLG theory \cite{Hosomichi:2008qk} which does not break any of the supersymmetries, though it breaks conformal invariance by the introduction of a mass parameter $m$. It also breaks $SO(8)$ R-symmetry to $SO(4)\times SO(4)$. The embedding of $SO(4)\times SO(4)$ in $SO(8)$ is such that $8_v \rightarrow 4_v + 4_v$. Accordingly we split the vector index $I$ as $(i,\hat{i})$. The mass deformed BLG supersymmetry variations are obtained by modifying the variation of the fermion by adding the term
\bea
\delta' \psi &=& m \Gamma \Gamma_I \epsilon X^I
\eea
Here 
\bea
\Gamma &=& \frac{1}{24} \epsilon^{ijkl}\Gamma_{ijkl}.
\eea
To maintain a maximally supersymmetric lagrangian, we add the following terms to the lagrangian \cite{Hosomichi:2008qk}
\bea
{\cal{L}} &=& - \frac{m^2}{2} \<X^I, X^I\> -\frac{im}{2}\<\bar{\psi}, \Gamma\psi\>\cr
&& - \frac{m}{6} \(\epsilon_{ijkl} \<X^i,\{X^j,X^k,X^l\}\> + \epsilon_{\hat{i}\hat{j}\hat{k}\hat{l}} \<X^{\hat{i}}\{X^{\hat{j}},X^{\hat{k}},X^{\hat{l}}\}\>\)
.
\eea
In a background with $\psi = 0$, the non-trivial condition for unbroken supersymmetry is that
\bea
\delta \psi &=& 0.
\eea
Assuming that only the four scalar fields $X^i$ are excited and $X^{\hat{i}} = 0$, the condition for unbroken supersymmetry, in a static field configuration, reads
\bea
0 &=& \(m X^i + \frac{1}{6} \epsilon_{ijkl} \{X^j,X^k,X^l\}\) \Gamma_i \epsilon
\eea
This condition does not restrict the supersymmetry parameter. We can write the condition for the maximally supersymmetric vacuum field configuration as
\bea
\{X^i,X^j,X^k\} &=& m \epsilon^{ijkl} X^l.
\eea
We can solve this equation by taking 
\bea
X^i X^i &=& \frac{1}{m^2}.
\eea
that is we find a three-sphere of radius
\bea
R &=& \frac{1}{m}.
\eea

\section{Constant spinor and the Nambu bracket}
In order to have closure of the BLG supersymmetry variations we must require that the supersymmetry parameter $\epsilon$ is such that
\bea
\{\epsilon,X^I,X^J\} &=& 0
\eea
This condition comes from taking a second supersymmetry variation on the fermion and demanding on-shell closure. Clearly we must extend our definition of the Nambu bracket to the case where the entries are not scalar entities. 

In ABJM theory with gauge group $U(N)\times U(N)$ say, for any finite $N$, apparently the supersymmetry parameter $\epsilon$ is just a constant,
\bea
\partial_M \epsilon &=& 0.
\eea
However this equation is not covariant, and is written in flat eleven-dimensional Minkowski coordinates $x^M$. We can write the condition in a covariant way as
\bea
D_M \epsilon = (\partial_M + \Omega_M)\epsilon = 0
\eea
where $\Omega_M$ is the spin connection. In the infinite $N$ limit we have a classical three-sphere and it is then more useful to express the constancy  condition in terms of the polar coordinates
\bea
x^M &=& x^M(x^{\mu},\theta^{\alpha},R,x^{\hat{i}})
\eea
for which the metric is given by
\bea
ds^2 &=& \eta_{\mu\nu}dx^{\mu}dx^{\nu}+g_{\alpha\beta}d\theta^{\alpha}d\theta^{\beta}+dR^2+dx^{\hat{i}}dx^{\hat{i}}
\eea
and we find the following non-vanishing Christoffel symbols,
\bea
\(A^T_{\alpha}\)^{\gamma}{}_{\beta} &=& \Gamma^{\gamma}_{\alpha\beta}\cr
\(A^N_{\alpha}\)^R{}_{\beta} &=& -\frac{g_{\alpha\beta}}{R}
\eea
which we interpret as two gauge fields associated with the tangent bundle and the normal bundle of the three-sphere respectively. In terms of these coordinates the Killing spinor equation pulled back to the three-sphere reads\footnote{In our conventions
\bea
D_{\alpha} &=& \partial_{\alpha} + \frac{1}{2}\Gamma_{a\alpha b} M^{ab}
\eea
where $a$ is a local flat index and the $SO(1,10)$ algebra generators $M^{ab}$ are normalized so that
\bea
[M_{ab},M^{cd}] &=& -4 \delta^{[c}_{[a} M^{b]}{}_{d]}
\eea
In vector and spinor representations we then find
\bea
(M^{ab})_{cd} &=& 2 \delta^{ab}_{cd},\cr
M^{ab} &=& \frac{1}{2}\Gamma_{ab}.
\eea
Here $\Gamma_{a\alpha b} = \eta_{ac}\Gamma^c_{\alpha b} = -\Gamma_{b \alpha a}$ is the Ricci rotation coefficient, or the Christoffel symbol with two indices converted into flat indices by means of two vielbeins.}
\bea
D_{\alpha} \epsilon = \(D_{\alpha}^T-\frac{1}{2R}\Gamma_R\Gamma_{\alpha}\)\epsilon = 0\label{psi}
\eea
where $D^T_{\alpha}\equiv \partial_{\alpha}+A^T_{\alpha}$ is the intrinsic covariant derivative on the three-sphere. Note that, since ${\mb{R}}^{1,10}$ is flat, we have 
\bea
[D_{\alpha},D_{\beta}] &=& 0.
\eea

The Killing spinor equation $D_{\alpha}\epsilon = 0$ means that we should define the Nambu bracket as
\bea
\{\epsilon,X^I,X^J\} &=& \sqrt{g}\epsilon^{\alpha\beta\gamma}D_{\alpha}\epsilon \partial_{\beta} X^I \partial_{\gamma} X^J.
\eea
This definition is crucial for getting closure of the $\N = 8$ supersymmetry. We note that
\bea
D_{\alpha}X^I = \frac{\partial X^I}{\partial \theta^{\alpha}} D_J X^I = \partial_{\alpha} X^I
\eea
upon taking the pull back to the three-sphere. It is a bit surprising that a covariant derivative can act on the $X^I$ just as if these were scalar fields since these do actually carry an R symmetry index and accordingly should rather be viewed as a section of an $SO(8)$ vector bundle over $S^3$. Let us therefore re-derive this `scalar' field property of the $X^I$ also in an intrinsic way, from the point of view of the three-sphere. On the three-sphere the only relevant non-vanishing Christoffel symbol $\Gamma^R_{\alpha\beta}$ couples to $X^{\beta}$ as $D_{\alpha}X^I = \partial_{\alpha}X^I + \Gamma^R_{\alpha\beta} X^{\beta}$. Since there is no field component $X^{\beta}$ in BLG theory we again conclude that $D_{\alpha}X^I = \partial_{\alpha}X^I$.

The general definition of the Nambu bracket must then be
\bea
\{f,g,h\} &=& *\(Df\wedge Dg \wedge Dh\)\label{Nambugeneral}
\eea
where $D$ denotes the covariant exterior derivative (including the normal bundle gauge field\footnote{I would like to thank Soo-Jong Rey for pointing out to me that one has to take into account the normal bundle gauge field}.). 

The real three-algebra is defined by a real three-bracket $[\cdot,\cdot,\cdot]$ satisfying the fundamental identity. The three-bracket is totally antisymmetric. We also require the existence of a positive definite trace form $\<\cdot,\cdot\>$ subject to the invariance condition
\bea
\<[T^c,T^d,T^a],T^b\> + \<T^a,[T^c,T^d,T^b]\> &=& 0.
\eea
The only finite-dimensional example is $SO(4)$. We also have infinite-dimensional algebras realized by the Nambu three-bracket. 

We define the associated trace form as
\bea
\<f,g\> &=& \int d^3 \theta \sqrt{g} fg.
\eea
We can expand any function in a complete basis of functions. We denote the basis elements by
\bea
T^a &=& T^a(\theta).
\eea
However, tensoring this basis element by an $\theta^{\alpha}$-independent spinor or tensor, it is essential to use the total covariant derivative $D_{\alpha}$ acting on the quantity. For instance we expand the BLG spinor in this three-algebra basis as
\bea
\psi(x,\theta) &=& \psi_a(x) T^a(\theta)
\eea
and compute its derivative as
\bea
D_{\alpha} \psi(x,\theta)
\eea
even though, of course $T^a$ is just a scalar entity, the $\psi_a(x)$ part carries R-symmetry indices associated both with space-time, internal three-manifold, and its normal bundle as embedded in eleven-dimensions. It is therefore essential that we use the total covariant derivative acting on $\psi(x,\theta)$. However, on the basis elements $T^a$ we act with the ordinary derivative $\partial_{\alpha} T^a$ since the basis functions are scalar quantities that carry no R symmetry indices nor spacetime indices.

We note that the fundamental identity  
\bea
\{T^a,T^{[b},\{T^c,T^d,T^{e]}\}\} &=& 0
\eea
is satisfied only if we can use ordinary commuting derivatives. We expand the left-hand side (defining $g\epsilon^{\alpha\beta\gamma}\epsilon^{\alpha'\beta'\gamma'} = 6g^{\alpha\beta\gamma,\alpha'\beta'\gamma} = g^{\alpha\alpha'}g^{\beta\beta'}g^{\gamma\gamma'}\pm $anti-symmetric.)
\bea
&& g \epsilon^{\alpha\beta\gamma} \epsilon^{\alpha'\beta'\gamma'} D_{\alpha} T^a D_{\beta} T^{[b} D_{\gamma} \(D_{\alpha'} T^c D_{\beta'} T^d D_{\gamma'} T^{e]}\)\cr
&=& 6D_{\gamma}\(g^{\alpha\beta\gamma,\alpha'\beta'\gamma'} D_{\alpha} T^a D_{\beta} T^{[b} D_{\alpha'} T^c D_{\beta'} T^d D_{\gamma'} T^{e]}\)\cr
&& - 6g^{\alpha\beta\gamma,\alpha'\beta'\gamma'} (D_{\gamma}D_{\alpha} T^a) D_{\beta} T^{[b} D_{\alpha'} T^c D_{\beta'} T^d D_{\gamma'} T^{e]}\cr
&& - 6g^{\alpha\beta\gamma,\alpha'\beta'\gamma'} D_{\alpha} T^a (D_{\gamma}D_{\beta} T^{[b}) D_{\alpha'} T^c D_{\beta'} T^d D_{\gamma'} T^{e]}.
\eea
We see that the last line vanishes only if the derivatives commute. 

We next check the trace invariance condition,
\bea
&&\<\{T^c,T^d,T^a\},T^b\> + \<T^a,\{T^c,T^d,T^b\}\> \cr
&=& \int d^3 \theta g \epsilon^{\alpha\beta\gamma} \(D_{\alpha}T^c D_{\beta}T^d D_{\gamma}T^a T^b + T^a D_{\alpha}T^b D_{\beta}T^c D_{\gamma}T^d\)\cr
&=& \int d^3 \theta g \epsilon^{\alpha\beta\gamma} D_{\alpha}T^c D_{\beta}T^d D_{\gamma}\(T^a T^b\) 
\eea
This vanishes only if we can write this as a total derivative. This will be the case in all cases we will be interested in. This is so because we use the trace form only to get the lagrangian. Since the lagrangian does not carry any indices we find that the total derivative is an ordinary derivative. Though if we act by an ordinary derivative on a contraction of two spinors for example, we find two covariant derivatives as
\bea
\partial_{\alpha} \(\bar{\psi}\psi\) &=& D_{\alpha}\bar{\psi} \psi + \bar{\psi} D_{\alpha}\psi.
\eea

Since 
\bea
D_{\alpha} X^I &=& \partial_{\alpha} X^I
\eea
we find that 
\bea
\{X^I,X^J,X^K\} &=& \sqrt{g}\epsilon^{\alpha\beta\gamma}\partial_{\alpha}X^I\partial_{\beta}X^J\partial_{\gamma}X^K
\eea
and we then find that the fundamental identity holds for these scalar fields,
\bea
\{\{X^{[I},X^J,X^K\},X^{L]},X^M\} &=& 0
\eea
This is enough to ensure supersymmetry.

\section{Computing the induced Lagrangian}
On the M5 brane we have a selfdual three-form which implies that there is no diffeomorphism invariant classical lagrangian formulation of the theory. However by giving up diffeomorphism invariance, we can find a lagrangian description. One example is given in \cite{Perry:1996mk} associated to the split of six dimensions into five plus one. Here we find a different version of such a diffeomorphism non-invariant lagrangian, associated to the split of six as three plus three. This lagrangian was also studied in \cite{Pasti:2009xc}.

We aim at finding a six dimensional lagrangian by expanding mass deformed BLG theory about the three-sphere vacuum. We want this six dimensional theory to possess as much diffeomorphism symmetry as possible.

\subsection*{Six-dimensional fluctuation fields}
The eight scalar fields correspond to fluctuations in eight dimensional transverse space. As we have already mentioned, we find it convenient to change coordinates as
\bea
x^I \mapsto x^I(\theta^{\alpha},x^A)
\eea
Then the three-sphere is a level curve, which we may choose to be located at $x^A = 0$,
\bea
T^I(\theta) &=& x^I(\theta,0)
\eea
We then consider small fluctuations of this three-sphere
\bea
\delta x^I(\theta,0) &=& \delta \theta^{\alpha} \partial_{\alpha} x^I(\theta,0) + \delta x^A \partial_A x^I(\theta,0) \label{fluct1}
\eea
For notational convenience, we define
\bea
Y^I(x,\theta) \equiv \delta x^I(x,\theta,0) \equiv X^I(x,\theta) - T^I(x,\theta)
\eea
where we re-instated the $x^{\mu}$ dependence as well, to illustrate that these are really six-dimensional fields. We associate six-dimensional fields to these fluctuations as
\bea
\delta \theta^{\alpha} &=& \phi^{\alpha}\cr
\delta x^A &=& \phi^A
\eea
As it turns out, the dual field $B_{\alpha\beta}$ defined as
\bea
\phi^{\alpha} &=& \frac{1}{2} \sqrt{g} \epsilon^{\alpha\beta\gamma} B_{\beta\gamma}
\eea
will be identified as components of a gauge potential in the M5 brane. 

We define the remaining gauge field components $B_{\mu\alpha}$ as
\bea
A_{\mu,ab}T^a \partial_{\alpha} T^b &=& B_{\mu\alpha}\label{fluct2}
\eea
It is not clear whether this relation can be inverted so as to express $A_{\mu,ab}$ in terms of $B_{\mu\alpha}$. Since our goal is to derive the M5 from M2 we will not need to invert this relation for our immediate purposes. However if we were to derive M2 from M5 it seems we would need to invert this relation.

We first show that $B_{\alpha\beta}$ and $B_{\mu\alpha}$ defined as above can really be identified as components of a two-form gauge potential in a six-dimensional theory, by showing that a gauge variation in BLG theory induces a gauge variation of these two-form components. A gauge transformation in BLG theory is given by
\bea
\delta X^I &=& \Lambda_{ab}(x) \{T^a,T^b,X^I\},\cr
\delta A_{\mu,ab} &=& D_{\mu} \Lambda_{ab}(x).
\eea
To linear order we find the induced gauge variations
\bea
\delta B_{\alpha\beta} &=& \partial_{\alpha}\Lambda_{\beta}-\partial_{\beta}\Lambda_{\alpha},\cr
\delta B_{\mu\alpha} &=& \partial_{\mu}\Lambda_{\alpha}-\partial_{\alpha}\Lambda_{\mu},\cr
\delta \phi^A &=& 0,
\eea
with gauge parameter
\bea
\Lambda_{\alpha} &=& \Lambda_{ab} T^a \partial_{\alpha} T^b,\cr
\Lambda_{\mu} &=& 0.
\eea

We note that
\bea
D_{\alpha}B_{\mu\beta} &=& D^T_{\alpha}B_{\mu\beta}
\eea
since there is no component $B_{\mu R}$. Also, then we have as usual that
\bea
D_{\alpha}B_{\mu\beta}-D_{\beta}B_{\mu\alpha} &=& \partial_{\alpha}B_{\mu\beta}-\partial_{\beta}B_{\mu\alpha}.
\eea
If we define
\bea
B_{\alpha\mu} &=& -B_{\mu\alpha}
\eea
then we define the field strength components as
\bea
H_{\mu\alpha\beta} &=& \partial_{\mu}B_{\alpha\beta} + \partial_{\alpha}B_{\beta\mu}-\partial_{\beta}B_{\alpha\mu},\cr
H_{\alpha\beta\gamma} &=& \partial_{\alpha}B_{\beta\gamma} + \partial_{\gamma}B_{\alpha\beta} + \partial_{\gamma}B_{\alpha\beta}.
\eea
To show that the action is supersymmetric we need Bianchi identities
\bea
D_{[\alpha} H_{\beta\gamma\delta]} &=& 0\cr
D_{[\alpha} H_{\mu\beta\gamma]} &=& 0\cr
D_{[\alpha} H_{\mu\nu\beta]} &=& 0
\eea
where we define
\bea
H_{\mu\nu\alpha} &=& \partial_{\mu}B_{\nu\alpha}-\partial_{\nu}B_{\mu\alpha}.
\eea

The supersymmetry parameter $\epsilon$ and spinor field $\psi$ in BLG theory are subject to chirality conditions
\bea
\t \Gamma \epsilon &=& \epsilon\cr
\t \Gamma \psi &=& -\psi.
\eea
These Weyl conditions are not six-dimensional. The `chirality matrix' associated with the three-manifold is given by
\bea
\Sigma &=& \frac{1}{6}\Gamma_{IJK}\{T^I,T^J,T^K\}\cr
&=& \frac{1}{6}\sqrt{g}\epsilon^{\alpha\beta\gamma} \Gamma_{\alpha\beta\gamma}.
\eea
This matrix has the anti-properties
\bea
\Sigma^{\dag} &=& -\Sigma,\cr
\Sigma^2 &=& -1
\eea
but when combined with the $SO(8)$ chirality matrix $\t \Gamma$, we find a true (six-dimensional) chirality matrix  
\bea
\t \Gamma \Sigma.
\eea
We would like to find new spinors $\omega$ and $\chi$ respectively, such that these are subject to the Weyl conditions
\bea
\t \Gamma \Sigma \omega &=& -\omega\cr
\t \Gamma \Sigma \chi &=& \chi
\eea
which are of a six-dimensional covariant form. We find these conditions by making the unitary rotation
\bea
\epsilon &=& U \omega\cr
\psi &=& U \chi\label{rot}
\eea 
with
\bea
U &=& \frac{i}{\sqrt{2}}\t \Gamma (1-\Sigma).\label{unitarym}
\eea

\subsubsection*{Scalar matter part}
The scalar matter field part is
\bea
{\cal{L}} &=& {\cal{L}}_{kin} + {\cal{L}}_{pot}
\eea
where
\bea
{\cal{L}}_{kin} &=& -\frac{1}{2}\<D^{\mu}X^I, D_{\mu}X^I\>\cr
{\cal{L}}_{pot} &=& -\frac{1}{12}\<\{X^I,X^J,X^K\},\{X^I,X^J,X^K\}\>-\frac{m^2}{2}\<X^I,X^I\>\cr
&&-\frac{m}{6}\epsilon^{ijkl}\<X^i\{X^j,X^k,X^l\}\>-\frac{m}{6}\epsilon^{\hat{i}\hat{j}\hat{k}\hat{l}}\<X^{\hat{i}}\{X^{\hat{j}},X^{\hat{k}},X^{\hat{l}}\}\>.
\eea

To zeroth order in fluctuation fields, we find that
\bea
{\cal{L}} &=& -\frac{1}{12}\<\{T^i,T^j,T^k\},\{T^i,T^j,T^k\}\> - \frac{m^2}{2}\<T^i,T^i\> - \frac{m}{6}\epsilon^{ijkl}\<T^i,\{T^j,T^k,T^l\}\>\cr
&=& -\frac{1}{2} - \frac{1}{2} + 1 = 0.
\eea
This being zero reflects the fact that the three-sphere solution is a supersymmetric ground state.

To first order we find
\bea
{\cal{L}} &=& -\frac{1}{2} \<\{T^i,T^j,T^k\},\{T^i,T^j,Y^k\}\> - m^2 \<T^i,Y^i\> - \frac{2m}{3} \epsilon^{ijkl} \<Y^i,\{T^j,T^k,T^l\}\>\cr
&=& (-3 + 4 - 1)m^2 \<T^i,Y^i\> = 0.
\eea
This being zero means that the three-sphere is a solution to the classical equation of motion. 

The first non-vanishing contributions starts at quadratic order. There will be higher order corrections but these are suppressed by an order of $1/R$ and can be ignored by taking $R$ sufficiently large. In this paper we will compute only up to quadratic order.

We start by computing the kinetic term. First we compute
\bea
D_{\mu} X^i &=& \frac{1}{2}\sqrt{g}\epsilon^{\alpha\beta\gamma}H_{\mu\alpha\beta}\partial_{\gamma}T^i+\partial_{\mu}\phi^R \partial_R T^i,\cr
D_{\mu} X^{\hat{i}} &=& \partial_{\mu}\phi^{\hat{i}}.
\eea
and consequently we get
\bea
{\cal{L}}_{kin} &=& -\frac{1}{4}H_{\mu\alpha\beta}H^{\mu\alpha\beta}-\frac{1}{2}\partial_{\mu}\phi^A\partial^{\mu}\phi^A.
\eea
We next expand the potential term,
\bea
{\cal{L}}_{pot} &=& -\frac{1}{2}\<\{Y^i,Y^j,T^k\},\{T^i,T^j,T^k\}\>\cr
&& -\frac{1}{2}\<\{T^i,T^j,Y^k\},\{Y^i,T^j,T^k\}\>\cr
&& -\frac{1}{4}\<\{T^i,T^j,Y^K\},\{T^i,T^j,Y^K\}\>\cr
&& -\frac{m^2}{2} \<Y^I,Y^I\> - m\epsilon^{ijkl}\<Y^i,\{Y^j,T^k,T^l\}\>.
\eea
We may use the trace invariance condition and the fundamental identity and get the identity
\bea
\<\{a,b,c\},\{e,f,g\}\> &=& 3 \<\{f,g,c\},\{e,a,b\}\>
\eea
where the right-hand side is to be antisymmetrized in $e,f,g$. Using this we can bring the lagrangian into the form of a sum of two terms,
\bea
{\cal{L}}_{pot} &=& {\cal{L}}_I + {\cal{L}}_m
\eea
where
\bea
{\cal{L}}_I &=& \frac{1}{2}\<\{T^j,T^k,Y^k\},\{T^j,T^i,Y^i\}\> - \frac{1}{4}\<\{T^i,T^j,Y^K\},\{T^i,T^j,Y^K\}\>\cr
{\cal{L}}_m &=& -\frac{m^2}{2} \<Y^I,Y^I\>.
\eea
We then note that\footnote{To see this we note that any vector in $\mb{R}^4$ can be written as
\bea
v^i &=& a T^i + b^{\alpha}\partial_{\alpha}T^i.
\eea
Then the identity can be proved by acting by both sides on this vector. We may also note the operator is a projector.}
\bea
g^{\alpha\beta} \partial_{\alpha} T^i \partial_{\beta} T^j &=& \delta^{ij} - \frac{T^iT^j}{R^2}
\eea
and we get
\bea
{\cal{L}}_I &=& -\frac{1}{2} \int d^3\theta \sqrt{g} \(g^{\gamma\gamma'}\frac{T^k T^{k'}}{R^2} + g^{\gamma\beta'}g^{\gamma'\beta}\partial_{\beta}T^k \partial_{\beta'}T^{k'}\)\partial_{\gamma}Y^k\partial_{\gamma'}Y^{k'}\cr
&&-\frac{1}{2}\int d^3\theta \sqrt{g} g^{\gamma\gamma'}\partial_{\gamma}Y^{\hat{k}}\partial_{\gamma'}Y^{\hat{k}},\cr
{\cal{L}}_m &=& -\frac{m^2}{2}\int d^3\theta \sqrt{g} \(Y^iY^i+Y^{\hat{i}}Y^{\hat{i}}\)
\eea

We now proceed by inserting the expansions in terms of fluctuation fields defined as in Eq (\ref{fluct1}), which we repeat here,
\bea
Y^i &=& \frac{T^i}{R}\phi^R+\phi^{\alpha}\partial_{\alpha}T^i,\cr
Y^{\hat{i}} &=& \phi^{\hat{i}}
\eea
From this it follows that
\bea
\partial_{\alpha}Y^i &=& \frac{1}{R}(\partial_{\alpha}T^i)\phi^R+\frac{T^i}{R}\partial_{\alpha}\phi^R+(D_{\alpha}^T\phi^{\beta})\partial_{\beta}T^i+\phi^{\beta}D_{\alpha}^T\partial_{\beta}T^i.
\eea
We have noted that $\partial_{\alpha}T^i$ transform as four vectors (one for each fixed value of $i$) on the three-sphere, or equivalently, that $\phi^{\alpha}\partial_{\alpha}T^i$ are four scalars on $S^3$. Consequently $\partial_{\alpha}(\phi^{\beta}\partial_{\beta}T^i) = (D_{\alpha}^T\phi^{\beta})\partial_{\beta}T^i + \phi^{\beta}D^T_{\alpha}\partial_{\beta}T^i$. We then note that\footnote{The left-hand side is symmetric. Hence one can suspect the result be proportional to $g_{\alpha\beta}$ (or to the Ricci tensor, but these are proportional on $S^3$). The normalization is then fixed by computing $T^i D_{\alpha}^T \partial_{\beta}T^i = - \partial_{\alpha}T^i \partial_{\beta}T^i = - g_{\alpha\beta}$ where we used the three-sphere constraint $T^i T^i = R^2$ to move one derivative.} 
\bea
D_{\alpha}^T\partial_{\beta}T^i &=& -\frac{1}{R^2}g_{\alpha\beta}T^i
\eea
on $S^3$.

The main point in this paper is to express everything in terms of total derivatives. We motivate this by the fact that the supersymmetry parameter is constant only with respect to the total derivative. By noting that the only non-vanishing Christoffel symbols in our polar coordinate system are 
\bea
\Gamma^{\gamma}_{\alpha\beta},\cr
\Gamma^R_{\alpha\beta} &=& - \frac{g_{\alpha\beta}}{R}\cr
\Gamma^{\alpha}_{\beta R} &=& \frac{1}{R}\delta^{\alpha}_{\beta}
\eea
We find that
\bea
D_{\alpha} \phi^{\beta} &=& D_{\alpha}^T \phi^{\beta} + \frac{1}{R} \delta_{\alpha}^{\beta} \phi^R,\cr
D_{\alpha} \phi^R &=& \partial_{\alpha} \phi^R - \frac{1}{R} \phi_{\alpha},\cr
D_{\alpha} \phi^{\hat{i}} &=& \partial_{\alpha} \phi^{\hat{i}}.
\eea
Using all this, we find that
\bea
\partial_{\alpha}Y^i &=& (D_{\alpha} \phi^{\beta}) \partial_{\beta} T^i + \frac{T^i}{R} D_{\alpha} \phi^R.
\eea
Inserting this into ${\cal{L}}_{pot}$ we find the result
\bea
{\cal{L}}_{pot} &=& -\frac{1}{2} D_{\alpha} \phi^{\beta} D_{\beta} \phi^{\alpha} - \frac{1}{2} g^{\alpha\beta} D_{\alpha} \phi^A D_{\beta} \phi^A\cr
&& - \frac{1}{2R^2} \phi^A \phi^A - \frac{1}{2R^2} g_{\alpha\beta} \phi^{\alpha}\phi^{\beta}.
\eea
The placements of the derivatives in the first term looks funny and a naive guess could be that this is something ugly and unwanted. But in fact this precise juxtaposition of the two derivatives turns out to be crucial for getting a gauge invariant action. We can not make integration by parts using $D_{\alpha}$ since it does not lead to a total derivative. We rather have that a total derivative (which vanishes upon integration over closed three-sphere) is given by
\bea
\int d^3 \theta \sqrt{g} D^T_{\alpha} V^{\alpha} &=& \int d^3 \theta \partial_{\alpha}(\sqrt{g} V^{\alpha}).
\eea
So we must express everything in terms of the intrinsic covariant derivative $D^T_{\alpha}$ before we can make integrations by parts. We then find
\bea
D_{\alpha} \phi^{\beta} D_{\beta} \phi^{\alpha} &=& D^T_{\alpha} \phi^{\alpha} D^T_{\beta} \phi^{\beta}  + \phi^{\beta}[D^T_{\alpha},D^T_{\beta}]\phi^{\alpha}\cr
&& + \frac{2}{R} \phi^R D_{\alpha}^T \phi^{\alpha} + \frac{3}{R^2} (\phi^R)^2
\eea
On a three-sphere of radius $R$ we have 
\bea
[D^T_{\alpha},D^T_{\beta}]\phi^{\alpha} &=& \frac{2}{R^2} \phi_{\beta}.
\eea
If then, we also expand out the other non-trivial term in ${\cal{L}}_{pot}$, which is
\bea
-\frac{1}{2} D_{\alpha}\phi^R D^{\alpha} \phi^R &=& - \frac{1}{2} D^T_{\alpha}\phi^R D^{T\alpha}\phi^R - \frac{1}{R} \phi^R D^T_{\alpha} \phi^{\alpha} - \frac{1}{2R^2} \phi^{\alpha}\phi_{\alpha}
\eea
then we find that the mass term $\phi^{\alpha}\phi_{\alpha}$ exactly cancels out in ${\cal{L}}_H$, and we end up with
\bea
{\cal{L}}_H &=& - \frac{1}{2}D^T_{\alpha} \phi^{\alpha} D^T_{\beta} \phi^{\beta} - \frac{1}{2}D^T_{\alpha}\phi^{A}D^{T\alpha}\phi^{A}\cr
&& - \frac{3}{2R} \phi^R D_{\alpha}^T \phi^{\alpha} - \frac{3}{2R^2} (\phi^R)^2 - \frac{1}{2R^2} \phi^A \phi^A
\eea

\subsubsection*{Chern-Simons term}
From the Chern-Simons term
\bea
{\cal{L}}_{CS} &=& \frac{1}{2}\epsilon^{\mu\nu\lambda}A_{\mu,ab}\partial_{\nu}A_{\lambda,cd}\<T^c,\{T^b,T^c,T^d\}\>
\eea
we get
\bea
{\cal{L}}_{CS} &=& \frac{1}{2}\epsilon^{\mu\nu\lambda}g\epsilon^{\alpha\beta\gamma}\partial_{\alpha}B_{\mu\beta}\partial_{\nu}B_{\lambda\gamma}
\eea

\subsubsection*{Fermionic part}
The fermionic part is
\bea
\frac{i}{2}\<\bar{\psi},\Gamma^{\mu}D_{\mu} \psi\>+\frac{i}{4}\<\bar{\psi},\Gamma_{ij} \{T^i,T^k,\psi\}\>-\frac{im}{2}\<\bar{\psi},\Sigma\Gamma_R\psi\>.
\eea
We expand the second term
\bea
\frac{i}{4}\<\bar{\psi},\Gamma_{ij} \{T^i,T^k,\psi\}\> &=& \frac{i}{2} \<\bar{\psi},\Sigma\Gamma^{\alpha}D_{\alpha}\psi\>.
\eea
We then make the field redefinition
\bea
\psi &=& U \chi,\cr
\bar{\psi} &=& \bar{\chi} V
\eea
and we get
\bea
\frac{i}{2}\<\bar{\chi},\Gamma^{\mu}\partial_{\mu}\chi\>+\frac{i}{2}\<\bar{\chi},\Gamma^{\alpha}D_{\alpha}\chi\>-\frac{i}{2R}\<\bar{\chi},\Sigma\Gamma_R\chi\>.
\eea
To get here we have used that
\bea
D_{\alpha}\Sigma &=& 0.
\eea

\subsection*{The induced Lagrangian}
Summing up all the various contributions, the resulting induced six dimensional action that we obtain up to quadratic order, is given by
\bea
S &=& \int d^3 x d^3 \theta \sqrt{g} \({\cal{L}}_H + {\cal{L}}_{\phi} + {\cal{L}}_{\psi}\)
\eea
where
\bea
{\cal{L}}_H &=& -\frac{1}{12} g^{\alpha\beta\gamma,\alpha'\beta'\gamma'}H_{\alpha\beta\gamma}H_{\alpha'\beta'\gamma'}-\frac{1}{4}\eta^{\mu\mu'}g^{\alpha\beta,\alpha'\beta'}H_{\mu\alpha\beta}H_{\mu'\alpha'\beta'}\cr
&&-\frac{1}{2} \epsilon^{\mu\nu\lambda}\epsilon^{\alpha\beta\gamma} \partial_{\beta} B_{\mu\alpha} \partial_{\nu} B_{\lambda \gamma}-\frac{1}{4R}\epsilon^{\alpha\beta\gamma}\phi^R H_{\alpha\beta\gamma},
\eea
\bea
{\cal{L}}_{\phi} &=& -\frac{1}{2}\(\partial_{\mu}\phi^A\partial^{\mu}\phi^A+g^{\alpha\beta}D^T_{\alpha}\phi^A D^T_{\beta}\phi^A\)\cr
&& - \frac{1}{2R^2} \phi^A \phi^A - \frac{3}{2R^2} \phi^R\phi^R
\eea
\bea
{\cal{L}}_{\psi} &=& \frac{i}{2}\bar{\chi}\Gamma^{\mu}\partial_{\mu}\chi+\frac{i}{2}\bar{\chi}\Gamma^{\alpha}D^T_{\alpha}\chi + \frac{i}{4R} \bar{\chi}\Sigma\Gamma_R \chi.
\eea
For ${\cal{L}}_{\psi}$ we have used 
\bea
D_{\alpha}\chi &=& D^T_{\alpha}\chi + \frac{1}{2R} \Gamma_R \Gamma_{\alpha} \Sigma \chi.
\eea
This follows from Eq (\ref{psi}) if one notes that $D_{\alpha}\Gamma_{\beta} = 0$ and $D_{\alpha}^T \Gamma_{\beta} = 0$. The first condition follows by requiring that $V_{\beta} = \bar{\psi}_1 \Gamma_{\beta} \psi_2$ transforms like a vector for any two BLG spinors $\psi_{1,2}$. The second condition can be seen by requiring 
\bea
D_{\alpha}V_{\beta} &=& D^T_{\alpha}V_{\beta}+\frac{1}{R}g_{\alpha\beta}V_R
\eea
If we assume that $D_{\alpha} \Gamma_{\beta} = 0$ then we get
\bea
D_{\alpha}V_{\beta} &=& (D_{\alpha} \bar{\psi}_1) \Gamma_{\beta} \psi_2 + \bar{\psi}_1 \Gamma_{\beta} D_{\alpha}\psi_2
\eea
We expand $D_{\alpha}\psi_{1,2} = D^T_{1,2}\psi_{1,2} - \frac{1}{2R}\Gamma_R \Gamma_{\alpha} \psi_{1,2}$, and we get
\bea
D_{\alpha}V_{\beta} &=& (D^T_{\alpha}\bar{\psi}_1)\Gamma_{\beta}\psi_2 + \bar{\psi}_1 \Gamma_{\beta} D^T_{\alpha}\psi_2 + \frac{1}{R}g_{\alpha\beta}V_R
\eea
and this equals $D^T_{\alpha}V_{\beta}+\frac{1}{R}g_{\alpha\beta}V_R$ only if 
\bea
D^T_{\alpha} \Gamma_{\beta} &=& 0.
\eea
Using this, and also 
\bea
D_{\alpha}g_{\beta\gamma} &=& 0,\cr
D^T_{\alpha}g_{\beta\gamma} &=& 0
\eea
which can be seen as a consequence of $D_{\alpha}\Gamma_{\beta} = 0 = D^T_{\alpha}\Gamma_{\beta}$, or it can be derived direcctly as $D_{\alpha}g_{\beta\gamma} = D^T_{\alpha}g_{\beta\gamma} + \Gamma_{\alpha\beta}^R g_{R\gamma} + .. = D^T_{\alpha} g_{\beta\gamma}$ since $g_{R\alpha} = 0$ by our choice of coordinates. Of course $D^T_{\alpha}g_{\beta\gamma} = 0$ is the familiar metric compatibility condition. Taken this together we conclude that 
\bea
D_{\alpha} U &=& 0,\cr
D_{\alpha}^T U &=& 0
\eea
where $U$ is defined as in Eq (\ref{unitarym}). we then get for any BLG spinor $\psi$ related to $\chi$ as $\chi = U\psi$, $\psi = -U\chi$,
\bea
D_{\alpha}\chi &=& U D_{\alpha} \psi\cr
&=& D^T_{\alpha}\chi+\frac{1}{2R}U\Gamma_R \Gamma_{\alpha} U\chi\cr
&=& D^T_{\alpha}\chi+\frac{1}{2R}\Gamma_R\Gamma_{\alpha}\Sigma \chi
\eea
as asserted.

We note that the equation of motion for the two-form $B_{\mu\alpha}$ becomes a total derivative \cite{Pasti:2009xc}. If then we vary $\partial_{\beta}B_{\mu\alpha}$, rather than $B_{\mu\alpha}$, then we find the equation of motion
\bea
\partial_{\mu}B_{\nu\alpha}-\partial_{\nu}B_{\mu\alpha} &=& -\frac{\sqrt{g}}{2}\epsilon_{\mu\nu\lambda}\epsilon_{\alpha\beta\gamma} H^{\lambda\beta\gamma}
\eea
This is the same equation of motion as we get directly from the BLG equation of motion Eq (\ref{BLG2}) by inserting our fluctuation expansion. To see this we first we contract Eq (\ref{BLG2}) by $T^a \partial_{\alpha} T^b$ and then insert the fluctuation field expansions into the resulting equation of motion. We also note the three-sphere constraint $T^i T^i = R^2$ which implies that $T^i D^{\lambda}\partial_{\alpha}T^i = -\partial_{\alpha}T^i D^{\lambda}T^i$.

The gauge field part of the lagrangian, ${\cal{L}}_H$, was also obtained in \cite{Ho:2008nn} and further studied in \cite{Pasti:2009xc}.

\section{Induced supersymmetry}
To get the supersymmetry variations we can expand the mass deformed BLG supersymmetry variations to linear order in the fluctuations. At zeroth order we have $\delta T^I = 0$. There are no `higher order' contributions to the variation $T^I$ since the higher order variations sit in the fluctuation fields $Y^I \equiv X^I-T^I$. At linear order we find the supersymmetry variations for the fluctuations as
\bea
\delta Y^I &=& i \bar{\epsilon} \Gamma^I \psi,\cr
\delta \psi &=& \Gamma^{\mu}\Gamma_I\epsilon \partial_{\mu}T^I+m\Sigma\Gamma_R\Gamma_I \epsilon T^I-\Sigma \epsilon\cr
&&-\Gamma^{\mu}\Gamma_I \epsilon A_{\mu,ab}\{T^a,T^b,T^I\}+\Gamma^{\mu}\Gamma_I \epsilon \partial_{\mu}Y^I\cr
&&-\frac{1}{2}\Gamma_{IJK} \epsilon \{Y^I,T^J,T^K\}\cr
&&+m\Sigma\Gamma_R\Gamma_I \epsilon Y^I,\cr
\delta A_{\mu,ab} &=& i \bar{\epsilon} \Gamma_{\mu}\Gamma_I T^I_{[a} \psi_{b]}
\eea
We can cancel the zeroth order contribution in $\delta \psi$ by taking $T^I$ to lie on a three-sphere of constant radius $R=1/m$. With this choice of radius we preserve maximal supersymmetry and the first line above in $\delta \psi$ vanishes. 

\subsubsection*{Supersymmetry variations of the Bosons}
From $\delta Y^I = \i\bar{\epsilon}\Gamma^i\psi$ we get
\bea
\delta \phi^A &=& i \bar{\omega} \Gamma^A \chi
\eea
and 
\bea
\delta B_{\alpha\beta} &=& i \bar{\omega} \Gamma_{\alpha\beta} \chi
\eea
and from $\delta A_{\mu,ab}$ we get
\bea
\delta B_{\mu\alpha} &=& i \bar{\omega} \Gamma_{\mu}\Gamma_{\alpha}\chi + \partial_{\alpha} \lambda_{\mu}
\eea
where
\bea
\lambda_{\mu} &=& \frac{i}{2} \bar{\epsilon} \Gamma_{\mu} \Gamma_R \psi
\eea
is a gauge parameter. We also note that 
\bea
\lambda_{\alpha} &=& \frac{i}{2} \bar{\epsilon} \Gamma_{\alpha} \Gamma_R \psi \equiv 0
\eea
so this is really a six-dimensional gauge parameter.

\subsubsection*{Supersymmetry variation of the Fermions}
We insert the expansion Eq (\ref{fluct1}) and Eq (\ref{fluct2}) and get
\bea
\delta \psi &=& \Gamma^{\mu}\Sigma\Gamma^{\alpha\beta}\epsilon \partial_{\alpha}B_{\mu\beta} + \Gamma^{\mu}\Gamma_{\alpha}\epsilon \partial_{\mu}\phi^{\alpha} - \Sigma \epsilon D_{\alpha} \phi^{\alpha}\cr
&& + \Gamma^{\mu}\Gamma_A \epsilon \partial_{\mu} \phi^A - \Gamma_A \Sigma \Gamma^{\alpha} \epsilon D_{\alpha} \phi^A\cr
&&+ \frac{1}{R} \Sigma \Gamma_R \Gamma_A \epsilon \phi^A - \frac{2}{R} \Sigma \Gamma_R \Gamma_{\alpha} \epsilon \phi^{\alpha} 
\eea
We then dualize $\phi^{\alpha}$ into $B_{\alpha\beta}$ and make a unitary rotation by means of the matrix $U$ to gain six-dimensional covariance. We then get 
\bea
\delta \chi &=& \frac{1}{2}\Gamma^{\mu}\Gamma^{\alpha\beta} H_{\mu\alpha\beta} + \frac{1}{6} \Gamma^{\alpha\beta\gamma}\omega H^D_{\alpha\beta\gamma}\cr
&&-\Gamma^{\mu}\Gamma_A \omega \partial_{\mu}\phi^A - \Gamma^{\alpha}\Gamma_{A}\omega D_{\alpha}\phi^A\cr
&&-\frac{1}{R}\Sigma \Gamma_R \Gamma_A \omega \phi^A + \frac{1}{2R} \Sigma\Gamma_R\Gamma^{\alpha\beta}\omega B_{\alpha\beta}.
\eea
where we introduced
\bea
H^D_{\alpha\beta\gamma} &=& D_{\alpha}B_{\beta\gamma} + D_{\gamma}B_{\alpha\beta} + D_{\beta}B_{\gamma\alpha}\cr
&=& H_{\alpha\beta\gamma} + \frac{1}{R}\sqrt{g}\epsilon_{\alpha\beta\gamma}\phi^R
\eea
In terms of $D^T_{\alpha}$ derivatives we then find the result
\bea
\delta \chi &=& \frac{1}{2}\Gamma^{\mu}\Gamma^{\alpha\beta} H_{\mu\alpha\beta} + \frac{1}{6} \Gamma^{\alpha\beta\gamma}\omega H_{\alpha\beta\gamma}\cr
&& -\Gamma^{\mu}\Gamma_A \omega \partial_{\mu}\phi^A - \Gamma^{\alpha}\Gamma_{A}\omega D^T_{\alpha}\phi^A \cr
&&-\frac{1}{R}\Sigma \Gamma_R \Gamma_A \omega \phi^A + \frac{1}{R} \Sigma \omega \phi^R . 
\eea
These supersymmetry variations must close on-shell on Lie derivatives on ${\mb{R}}^{1,2}\times S^3$, the $SO(4)\subset SO(5)$ R-symmetry that keeps $\phi^R$ fixed, and a gauge variation as these are the bosonic symmetries of the action.

\section{Open problems}

By taking $k$ large we reduce $S^3$ to $S^2$ by shrinking the Hopf circle $k$ times due to the ${\mb{Z}}_k$ orbifold identification, and the M5 brane wrapped on $S^3$ reduces to D4 wrapped on $S^2$. In would be interesting to demonstrate this explicitly in our abelian theory and make connection to \cite{Nastase:2009ny}. Also since we know the nonabelian D4 brane theory this can give a hint of the nonabelian M5 brane theory.

One may consider more general mass deformations that still preserve maximal supersymmetry \cite{Lambert:2009qw}. It would be interesting to see what M5 brane theories these correspond to. We may also get less supersymmetric six-dimensional theories by expanding BLG theory about less supersymmetric backgrounds, such as has been classified in \cite{Jeon:2008zj}. In particular one can consider the half BPS funnel solution \cite{Basu:2004ed} of M2's ending on M5 and find a six dimensional theory with eight supercharges on curved manifold of the geometry of a funnel.

The right way to discretize the BLG theory with a Nambu bracket should be to consider ABJM theory. Needless to say it will be very interesting to derive BLG theory on $S^3/{\mb{Z}_k}$ by taking the large $N$ limit of mass deformed ABJM theory at level $k$. For $k=1,2$ we can indeed see the fuzzy three-sphere (mod ${\mb{Z}}_{2}$) in ABJM theory. Only for levels $k=1,2$ do we have enhanced $SO(8)$ R symmetry in ABJM theory for generic gauge groups. In this case we can find a fuzzy funnel solution that is locally a fuzzy three-sphere. We have not yet verified that a similar type of enhancement works also for the mass deformed ABJM theory eventhough this seems very plausible, so let us demonstrate how the fuzzy funnel solution arises. For levels $k=1,2$ we have showed in \cite{Gustavsson:2009pm} that the supersymmetry variation of the fermion in ABJM theory can be written as (using the same notations as in that paper)
\bea
\delta \psi &=& \Gamma^{\mu}\Gamma_I \epsilon D_{\mu} X^I - \frac{1}{6}\Gamma_I \Gamma_J \Gamma_K \epsilon [X^I,X^J;X^K].
\eea
Moreover we can antisymmetrize $IJK$ despite the three-bracket is only manifestly antisymmetric in its first two entries. This follows from the identity
\bea
X^I_b X^J_c X^{Kd} f^{bc}{}_{da} &=& X^K_b X^{[I}_c X^{J]d} f^{bc}{}_{da}
\eea
We then find that
\bea
\Gamma_{IJ}[X^I,X^J;X^K]_a &\equiv & \Gamma_{IJ} X^I_b X^J_c X^{Kd} f^{bc}{}_{da}\cr &=& \Gamma_{IJ} X^K_b X^{I}_c X^{Jd} f^{bc}{}_{da} \cr
&=& -\Gamma_{IJ} X^I_b X^K_c X^{Jd} f^{bc}{}_{da}
\eea
Hence the bracket can be antisymmetrized in $I,J,K$ when contracted by $\Gamma_{IJ}$. From here we can then derive the Basu-Harvey fuzzy three-sphere funnel solution \cite{Basu:2004ed} by requiring $\delta \psi = 0$. For level $k=2$ the $\mb{Z}_2$ orbifolding is just $X^I \sim -X^I$ that comes from $Z^A_a \sim -Z^A_a = e^{i\pi} Z^A_a$. The relation between $X^I_a$ and $Z^A_a$ involves a Wilson line $W_{ab}$ and the higher $k$ orbifolding $Z^A_a \sim e^{2\pi i/k} Z^A_a$ has no such simple counterpart for the $X^I_a$. Also the Wilson line becomes non-local and it is unclear to us whether one can find a fuzzy three-sphere mod $\mb{Z}_k$ also for higher levels $k$.

In \cite{Park:2008qe} it was demonstrated how the Nambu-Goto action for a five-brane  can be reformulated as a BLG type of theory with a Nambu three-bracket. It will be interesting to generalize this approach to the full-fledged kappa symmetric M5 brane action \cite{Bandos:1997ui} and derive (mass deformed) BLG theory from this action.

In principle the theory of multiple M5 branes should also be encoded in some ABJM theory. It would be very interesting to see if one can compute any quantity in the multiple M5 brane theory from ABJM theory. For finite rank gauge groups we would expect to find a non-commutative, and perhaps also non-abelian, M5 brane.

\vskip40pt
{\sl{Acknowledgements}:} I would like to thank Soo-Jong Rey for initiating this project and for discussions. I would like to thank Bum-Hoon Lee, Jeong-Hyuck Park and Takao Suyama for discussions and Neil Lambert for correspondence. This work was supported by Center for Quantum Spacetime of Sogang University with grant number R11-2005-021.

\newpage
\appendix
\section{Gamma matrix relations}
For the matrices 
\bea
U &=& \frac{i}{\sqrt{2}} \t \Gamma (1-\Sigma),\cr
V &=& -\frac{i}{\sqrt{2}} (1-\Sigma) \t \Gamma.
\eea
we have used the following identities,
\bea
U \Sigma \Gamma_R \Gamma_{\alpha} U &=& \Gamma_R \Gamma_{\alpha}\cr
U \Gamma^{\mu} \Sigma \Gamma^{\alpha\beta} U &=& \Gamma^{\mu}\Gamma^{\alpha\beta}\cr
U \Gamma^{\alpha\beta\gamma} U &=& \Gamma^{\alpha\beta\gamma}\cr
U \Gamma^{\mu} \Gamma_A U &=& \Gamma^{\mu} \Gamma_A\cr
U \Gamma^{\alpha} \Gamma_A \Sigma U &=& \Gamma^{\alpha} \Gamma_A\cr
U \Sigma U &=& \Sigma.\label{gamma1}
\eea
and
\bea
V\Gamma^{\mu} U &=& \Gamma^{\mu},\cr
V\Sigma \Gamma^{\alpha} U &=& \Gamma^{\alpha},\cr
V \Gamma_{\mu}\Gamma_{\alpha} U &=& - \Gamma_{\mu}\Gamma_{\alpha},\cr
V \Gamma_{A} U &=& -\Gamma_{A},\cr
V \Gamma_{\alpha} U &=& \Gamma_{\alpha}\Sigma\cr
V \Sigma \Gamma_{\alpha\beta} U &=& -\Gamma_{\alpha\beta}.\label{gamma2}
\eea

Our gamma matrices are subject to the algebra
\bea
\{\Sigma,\t \Gamma\} &=& 0,\cr
\{\Gamma_{\mu},\Sigma\} &=& 0,\cr
[\Gamma_{\alpha},\Sigma] &=& 0,\cr
\{\Gamma_A,\Sigma\} &=& 0,\cr
[\Gamma_{\mu},\t \Gamma] &=& 0,\cr
\{\Gamma_{\alpha},\t \Gamma\} &=& 0,\cr
\{\Gamma_A,\t \Gamma\} &=& 0,\cr
\{\Gamma_{\mu},\Gamma_{\alpha}\} &=& 0,\cr
\{\Gamma_{\mu},\Gamma_A\} &=& 0,\cr
\{\Gamma_{\alpha},\Gamma_A\} &=& 0.
\eea
and duality relations
\bea
\Sigma \Gamma^{\gamma} &=& \frac{1}{2} \sqrt{g} \epsilon^{\alpha\beta\gamma} \Gamma_{\alpha\beta},\cr
\Gamma^{\gamma} &=& -\frac{1}{2} \sqrt{g} \epsilon^{\alpha\beta\gamma}\Sigma \Gamma_{\alpha\beta},\cr
\Gamma_{\gamma} &=& -\frac{1}{2}\sqrt{g} \epsilon_{\alpha\beta\gamma} \Sigma \Gamma^{\alpha\beta},\cr
\Gamma_{\gamma} \epsilon^{\alpha\beta\gamma} &=& -\frac{1}{\sqrt{g}} \Sigma \Gamma^{\alpha\beta}.
\eea

\end{document}